\documentclass[]{interact}

\usepackage{epstopdf}
\usepackage[caption=false]{subfig}

\usepackage{algorithm}
\usepackage{algpseudocode}
\usepackage[inline]{enumitem}
\usepackage{mathtools}
\usepackage[draft]{todonotes}
\usepackage{tabularx}

\usepackage[natbibapa,nodoi]{apacite}
\usepackage[colorlinks=true, citecolor=blue,allcolors=blue]{hyperref}%

\theoremstyle{plain}

\theoremstyle{definition}

\theoremstyle{remark}

\usepackage{graphicx}
\usepackage{amsmath}
\usepackage{amsfonts}
\usepackage{mathrsfs}
\usepackage[OT1]{fontenc} 
\usepackage{enumerate}
\usepackage{amsthm}
\usepackage{tabularx}
\usepackage{arydshln}

\usepackage{subfig}
\usepackage{float}
\usepackage{graphicx}
\usepackage{textcomp}
\usepackage{xcolor}
\usepackage{footnote}
\usepackage{makecell}
\usepackage[draft]{todonotes}

\usepackage{algorithm}
\usepackage{algpseudocode}
\usepackage[inline]{enumitem}
\usepackage{mathtools}

\begin{document}

\articletype{ }

\title{Unconventional Arterial Intersection Designs under Connected and Automated Vehicle Environment: A Survey}

\author{
\name{Zijia Zhong, Mark Nejad, and Earl E. Lee}
\affil{ Department of Civil \& Environmental Engineering, University of Delaware, Newark, DE, United States }
}

\maketitle

\begin{abstract}
Signalized intersections are major sources of traffic delay and collision within the modern transportation system.  Conventional signal optimization has revealed its limitation in improving the mobility and safety of an intersection. Unconventional arterial intersection designs (UAIDs) are able to improve the performance of an intersection by reducing phases of a signal cycle.  Furthermore, they can fundamentally alter the number and the nature of the conflicting points.  However, the driver’s confusion, as a result of the unconventional geometric designs, remains one of the major barriers for the widespread adoption of UAIDs. Connected and Automated Vehicle (CAV) technology has the potential to overcome this barrier by eliminating the driver’s confusion of a UAID. Therefore, UAIDs can play a significant role in transportation networks in the near future. In this paper, we surveyed UAID studies and implementations. In addition, we present an overview of intersection control schemes with the emergence of CAV and highlight the opportunity rises for UAID with the CAV technology. It is believed that the benefits gained from deploying UAIDs in conjunction with CAV are significant during the initial rollout of CAV under low market penetration.

\end{abstract}

\begin{keywords}
Connected and Automated Vehicle; Unconventional Arterial Intersection Design; Driver’s Confusion; Intersection Traffic Management 
\end{keywords}

\section{Introduction}
Traffic engineers have been challenged to develop solutions to mitigate congestion, enhance safety, and improve the level of service (LOS) of at-grade intersections. Thus far, the solutions can be categorized into three groups: 1) traditional/conventional measure targeting on improving Signal Phase and Timing (SPaT) plan via coordination or optimization 2) grade-separated re-configuration of intersection geometry; and 3) deploying unconventional arterial intersection designs (UAIDs). 

The traditional approach via SPaT optimization is no longer able to considerably alleviate congestion at signalized intersections \citep{dhatrak2010performance}. The premise of signal optimization is that if the critical lane groups get served with longer phases, the capacity of an intersection could be increased. Therefore, adaptive signal control technologies (ASCTs) (e.g., ACS-Lite, SCATS, SCOOT) have been deployed as the traditional approach to provide adjustments on SPaT according to real-time traffic information. However, ASCTs do not change the geometry of the intersection and consequently, the conflict points of the intersection remain the same. 

The capacity of an intersection is the summation of all the lane groups. Based on this premise, the intersection capacity increases if more lane groups are served simultaneously. Grade separation can achieve this very objective, resulting in uninterrupted traffic flows.  It increases intersection safety at the same time as well. For instance, overpasses can significantly increase the capacity of an intersection by making approaches bypassing each other. Additionally, fewer signal phases are needed for directing traffic at a grade-separated intersection. However, the grade-separation approach often comes with a high cost of infrastructure investment. Only under few circumstances is the conversion economically justified.

UAIDs, also known as alternative intersection designs (AIDs), have the potential in improving the efficiency and safety of an intersection by strategically eliminating or changing the nature of the intersection conflict points. Various UAIDs have been proposed along the years (Table \ref{tbl:US_deployment}), but only a handful of them are seen in real-world deployments. The adoption of UAIDs in the US as of 2018 is shown in Figure 1. While the adoption of UAIDs exhibits an increasing trend in the US, additional research for UAID is still needed. Besides, little research has been done to UAIDs in the context of CAV technology. Thus far, numerous studies have been presented for the vision of full market penetration of CAVs where signalized intersections have become obsolete. However, in the near future, mixed traffic conditions would be the reality. Hence the deployment and promotion of CAVs still have to resort to the existing roadway infrastructure.  One of the major concerns for UAID is driver's confusion. It can be greatly mitigated or even eliminated with the aid of CAV, usually in the form of an advanced driver-assistance system (ADAS), which is typically referred as the lower levels of automation (below SAE Level 3 vehicle automation).

\begin{figure} [h]
	\centering
	\label{fig:location}
	\includegraphics[width=0.8\textwidth]{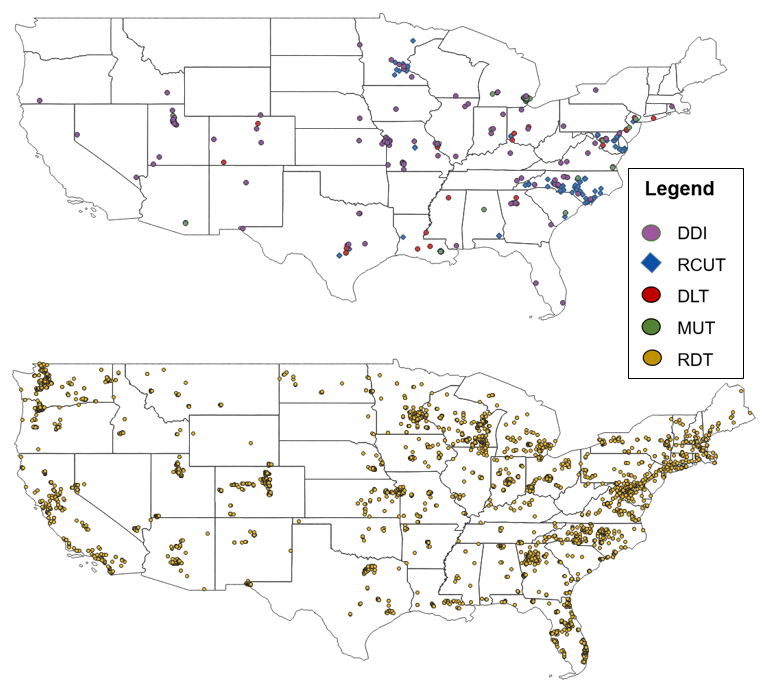}   
	\caption{UAID Deployment in the contiguous US (data source: \cite{InstitueforTransportationResearchandEducation}} 	
\end{figure}

\begin{table}[!h]
\centering
\caption{Year of UAID Introduction} 
\resizebox{0.9\textwidth}{!}
{
\begin{tabular}{p{3in}|p{2in}} \hline 
\textbf{UAID} & \textbf{Year Introduced} \\ \hline 
Roundabout*  & unknown  \\ \hline 
Median U-turn intersection (MUT)* & 1920s \citep{bessert2011michigan} \\ \hline 
Jughandle* & unknown  \\ \hline 
Bowtie & 1995 \citep{hummer2016safety} \\ \hline 
Split-phasing intersection (SPI) & 2003 \citep{chlewicki2003new}\\ \hline 
Diverging diamond interchange (DDI)* & 2003 \citep{chlewicki2003new}  \\ \hline 
Upstream signalized crossover (USC) & 2005 \citep{tabernero2006upstream} \\ \hline 
Center turn overpass (CTO) & 2005 \citep{Reid2004} \\ \hline 
Parallel flow intersection (PFI) & 2007 \citep{parsons2009parallel} \\ \hline 
Restricted crossing U-turn/ Superstreet intersection (RCUT)* & 2009 \citep{shahi2009modelling}\newline  \\ \hline 
Triangabout  & 2014 \citep{Chou2014}\\ \hline 
Symmetric intersection  & 2018 \citep{li2018symmetric} \\ \hline
Super Diverging Diamond Interchange  & 2019 \citep{molan2018introducing} \\ \hline
\end{tabular}
}
\label{tbl:US_deployment}
\end{table}

This paper aims to review the state of the research for UAIDs and the promising improvement which are made possible with emerging technology.  The remainder of the paper is organized as follows. Section \ref{UAIDDesign} summarizes the previous studies regarding UAIDs. Section \ref{sect: signalizedCAV} reviews the signalized intersection control schemes under CAV environment.  In Section \ref{sect: signalFreeCAV}, the signal-free intersection control for UAID is covered. We discuss and highlight the research trends for future UAID research under CAV environment in Section \ref{sect: Discussion}.  Lastly, we conclude the paper with final remarks in Section \ref{sect: conclusion}.

\section{Unconventional Arterial Intersection Designs}
\label{UAIDDesign}
\begin{figure} [h]
	\centering
	\label{fig:CP_Std}
	\includegraphics[width=0.8\textwidth]{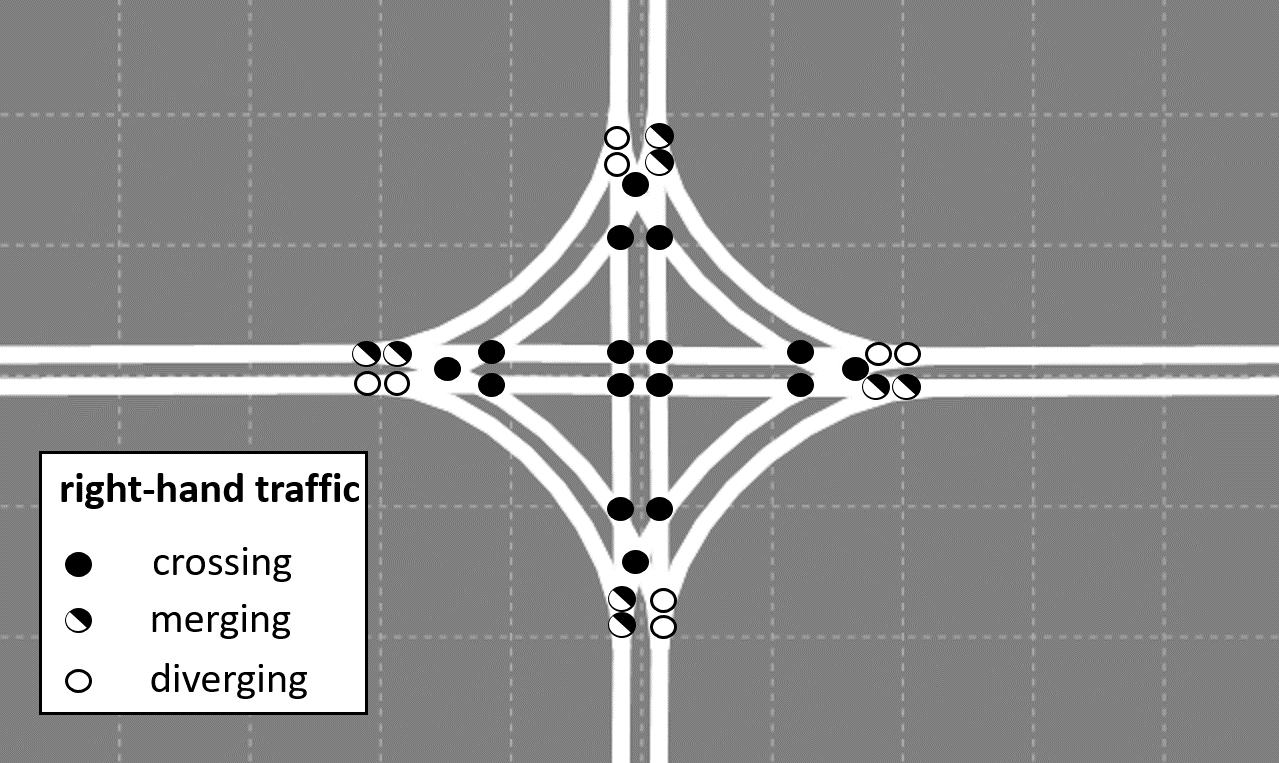}   
	\caption{Conflict points of a four-leg intersection} 	
\end{figure}

There are two underlying design concepts for UAIDs. The first one is to facilitate through traffic movements. The second one is to reduce the conflicts between left-turn movements and opposing through movements by re-routing some lane groups. In a standard four-leg intersection with four signal phases, there are 12 movements (four left-turning, four through, and four right-turning), which form 16 conflicting points as shown in Figure \ref{fig:CP_Std}. Among them, 12 of the conflict points are caused by the left turning movements. Not only does the left-turn movements hinder intersection performance, but they also attribute to most of the right-angle collisions, causing serious injuries for motorists and pedestrians.  

Table \ref{table:conflict_pt} shows the comparisons of common UAIDs with a standard four-leg intersection in terms of the number of the type of conflict points. As seen, the aforementioned four UAIDs all have fewer conflicts points than a standard intersection.

\begin{table}[!h]
\centering
\caption{Intersection conflict points} 
\resizebox{0.9\textwidth}{!}
{
\begin{tabular}{c|c|c|c|c} \hline 
\textbf{UAID} & \textbf{Conflict Pt. } & \textbf{Merging Pt.} & \textbf{Diverging Pt.} & \textbf{Crossing Pt.} \\ \hline 
\textbf{conventional} & 32 & 8 & 8 & 16 \\ \hline 
\textbf{DDI} & 14 & 6 & 6 & 12 \\ \hline 
\textbf{DLT} & 28(full)/\newline 30 (half) & 8 & 8 & 12(full)/\newline 14(half) \\ \hline 
\textbf{RCUT} & 8 & 4 & 4 & 0 \\ \hline 
\textbf{roundabout} & 8 & 4 & 4 & 0 \\ \hline 
\end{tabular}
}
\label{table:conflict_pt}
\end{table}

\subsection{UAID Designs}
In the Every Day Count initiative \citep{EDCSummit2013}, the Federal Highway Administration (FHWA) recommends the use of DDI, DLT, MUT (e.g., restricted crossing U-turns, median U-turns, and through U-turns), and roundabout. Considering the current the field deployment,  The characteristics of the four aforementioned UAIDs are discussed in the following subsections.   
\subsubsection{Diverging Diamond Interchange (DDI)}
As illustrated in Figure \ref{fig:geo_condition}(a), the concept of DDI is to eliminate the necessity of left-turn bays and accompanying signal phases at signalized ramp terminals between an arterial and a freeway. The upstream signalized crossovers operate with two-phase signal control.  Fewer conflict points are with DDI, which is likely leading to fewer crashes \citep{maze2004rural}. However, the counter-intuitive travel between the ramp terminals and the interchanges, and the crossing over to the other side of the traffic on the bridge may create confusion for drivers. The first DDI in the US was constructed at the crossing of I-44 and US-13 in Springfield, MO \citep{afshar2009traffic}.

\subsubsection{Displaced Left-turn Intersection (DLT)}
The DLT eliminates the conflict between left-turn and opposing through movements by displacing the left-turn lane to the opposing direction at upstream of the primary intersection as demonstrated in Figure \ref{fig:geo_condition}(b) Compared to grade-separated interchange, DLT can be constructed much faster and with less cost \citep{HermanusSteyn2014}. 
\subsubsection{Median U-turn Intersection (MUT)}
A MUT intersection aims to eliminate direct left turns from major and minor approaches by re-routing the traffic via a median located on the major street. As shown in Figure \ref{fig:geo_condition}(c), a driver has to drive past the intersection that he or she is intended to make a left turn, make a U-turn at the median opening, and then make a right-turn to complete the whole left-turn maneuver. The MUT can be implemented as RCUT (Restricted Crossing U-turn intersection) which prohibits through and left movements of the traffic from the side street. Similarly, drivers on the side street must first take a right turn, then a U-turn a left turn on the arterial to complete the through movement.
\subsubsection{Roundabout}
\begin{figure}[h]
\begin{minipage}[h]{.5\linewidth}
\centering
\subfloat[Diverging Diamond Interchange]{\label{main:a}\includegraphics[scale=.4]{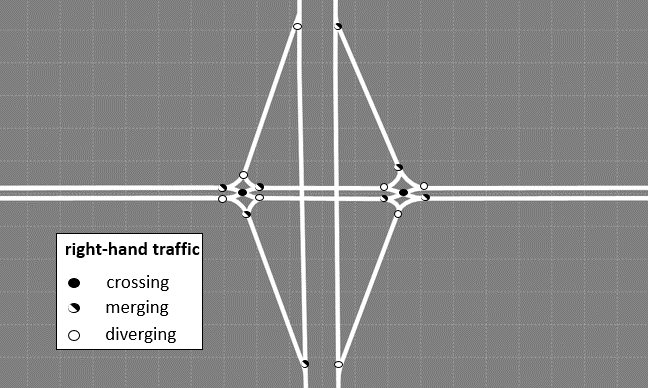}}
\end{minipage}%
\begin{minipage}[h]{.5\linewidth}
\centering
\subfloat[Displaced Left-Turn Intersection]{\label{main:b}\includegraphics[scale=.3]{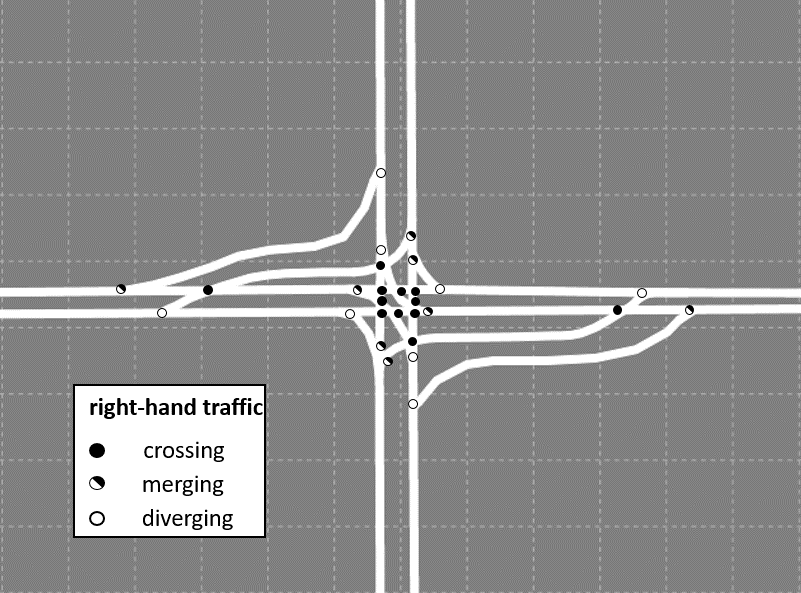}}
\end{minipage}\par
\begin{minipage}[h]{.5\linewidth}
\centering
\subfloat[Restricuted Median U-Turn Intersection]{\label{main:c}\includegraphics[scale=.35]{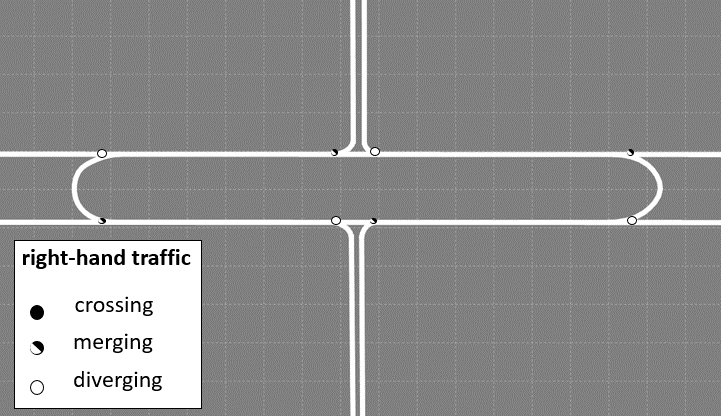}}
\end{minipage}%
\begin{minipage}[h]{.5\linewidth}
\centering
\subfloat[Roundabout]{\label{main:d}\includegraphics[scale=.5]{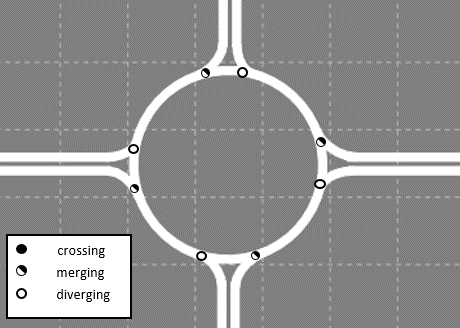}}
\end{minipage}\par
\caption{Geometry of Common UAIDs}
\label{fig:geo_condition}
\end{figure}

A roundabout changes the type of crashes at an intersection. roundabouts have consistently gained recognition as a safe alternative to tradition four-leg intersection in the US.  As demonstrated in Figure \ref{fig:geo_condition}(d), traffic enters and exits a roundabout only through making right turns and then proceeding in a counter-clockwise pattern. The traffic inside the roundabout has the right-of-way (ROW) over the incoming traffic. For a multi-lane roundabout, ideally, a driver should know the lane that he or she wishes to go prior to entering the roundabout. As of Jun. 2018, there are approximately 3,200 roundabouts in the US. France is the leading adopter nations for roundabouts with 30,000 roundabouts nationwide \citep{badgley2018fhwa}.

\subsection{Empirical Study}

Only limited empirical studies of UAIDs have been conducted, primarily due to the limited deployments in the real world. The roundabout is the only UAID that has a capacity model in the Highway Capacity Manual \citep{rodegerdts2015accelerating}. The empirical capacity model for the US roundabout can be found on the NCHRP Report 572 \citep{rodegerdts2007national}. \cite{isebrands2009crash} conducted an empirical study regarding the effectiveness of 17 existing high-speed (40+ mph) roundabouts in rural areas in the US. The analysis revealed an 84\% reduction in the crash frequency, an 89\% reduction the average injury crash rate, an 86\% reduction in angle crashes, and a 100\% reduction in fatal crashes . 

Safety evaluation for the seven DDIs deployed in the US was performed by \cite{hummer2016safety}. They aimed to recommend the crash modification factor (CMF) for converting to a DDI from a conventional diamond interchange (CDI) based on crash data.  The crash frequency, crash type, and traffic volume were taken into consideration. The CMF for the number of crashes was found to be 0.67, which means 33\% reduction in the total number of crashes. They also found the CMF for injury crashes to be 0.59.  A before-and-after study was conducted for MUTs, DLTs, and roundabouts in the State of Maryland \citep{kim2007unconventional}.  In this study, \cite{Claros2017} investigated the crash reports for nine DDIs and five MUTs .  It was found a 62.6\% reduction in fatal and injury crashes, 35.1\% reduction on property-damage-only crashes, and 47.9\% reduction on the total crashes for the DDIs. For the MUTs, the injury crashes and total crashes were reduced by 63.8\% and 31.2\%, respectively.  It was also revealed that the left-turn angle crashes were substantially decreased with the adoption of UAIDs.

The analytical approaches for studying UAIDs have been reported. \cite{yang2014development} developed a signal optimization model for DDI, in which the common cycle length and green split for crossover intersections were determined and the adjacent conventional intersections of the DDI were taken into account. A two-stage solution was proposed. The first stage aims to maximize the traffic throughput by optimal green slits and the second stage was to optimize the offsets (between primary and secondary intersections) by a modified Maxband method \citep{little1981maxband}.  \cite{Chlewicki2011} compared DDI, CDI, and SPUI by using the critical lane volume (CLV) method. He concluded that a DDI is not always the best option, but DDI has better traffic operation most of the time when the costs are similar.  The more lanes are needed, the more likely DDI will be a  better option.

\subsection{Simulation Study of UAID}

Besides empirical study, traffic simulation is the most common method for researching UAIDs. Commercial-of-the-shelf microscopic traffic simulation tools, including Vissim, CORSIM, PARAMICS, SimTraffic, AIMSUN, were often used.  Recent studies of UAID using simulation software are listed in Table \ref{table: uaidType}.  The studies are categorized into two groups:  the group using empirical data to calibrate the network and the group without it.

\begin{table}[h]
\centering
\caption{UAID Studied} 
\resizebox{1\textwidth}{!}
{
\begin{tabular}{@{}p{4cm}ccccccccccccccc@{}}
\toprule
\multicolumn{1}{c}{{Study}} & \multicolumn{2}{l}{Data Type} & \multicolumn{9}{c}{UAID}  \\ \cline{2-16} 
\multicolumn{1}{c}{}  & 
\multicolumn{1}{c}{E} & \multicolumn{1}{c}{S} & 
\multicolumn{1}{c}{DDI} & \multicolumn{1}{c}{MUT} & \multicolumn{1}{c}{DLT} & \multicolumn{1}{c}{RDT} & \multicolumn{1}{c}{USC} & \multicolumn{1}{c}{QR} & \multicolumn{1}{c}{PFI} 
& \multicolumn{1}{c}{DXI} & \multicolumn{1}{c}{SPI} & \multicolumn{1}{c}{JUG}& \multicolumn{1}{c}{Other} \\ \hline
 \cite{Warchol2017}* &  \checkmark &  & \checkmark & & &  & & &  & & &
 \\
 \cite{elazzony2010microsimulation}* &  \checkmark &  &  & \checkmark& &  & & & & & &
 \\
 \cite{esawey2011unconventional}* &  \checkmark &  &  & & &  & \checkmark & & & & &
 \\
 \cite{Chou2014}* &  \checkmark &  & \checkmark & & &  & & &  & & & &TDT
 \\
 \cite{chilukuri2011diverging}* &  \checkmark &  & \checkmark & & &  & & &  & &&
 \\
 \cite{chlewicki2003new} &  \checkmark &  & \checkmark & & &  & & &  & &\checkmark& &
 \\
 \cite{yang2014development} &  \checkmark &  & \checkmark & & &  & & &  & &
 \\
 \cite{Reid2001} &  \checkmark &  &  & \checkmark  & &\checkmark   & & \checkmark &  & &\checkmark &\checkmark& SSM, CFL
 \\
 \cite{el2011operational} &  \checkmark &  &  &\checkmark & &  & \checkmark& &  & \checkmark&&
 \\
 \cite{hughes2010alternative} &  \checkmark &  &  &\checkmark & \checkmark&  & &\checkmark &  && && DCD
 \\
 \cite{Tarko2017} &   & \checkmark & \checkmark  & & & \checkmark  & & & &&\checkmark && CDI, TDI
 \\
 \cite{Park2017} &   & \checkmark &\checkmark  & & &  & & &  & &
 \\
 \cite{autey2013operational} &   & \checkmark &  &\checkmark & \checkmark&  &\checkmark  & & &\checkmark&&
 \\
 \cite{edara2005diverging} &   & \checkmark &\checkmark  & & &  & & &   &\checkmark&&
 \\
 \cite{cheong2008comparison} &   & \checkmark &  & &\checkmark &  & \checkmark& &  \checkmark& &&&
 \\
 \cite{dhatrak2010performance} &   & \checkmark &  & & \checkmark&  & & &  \checkmark& &&&
 \\
 \cite{gallelli2008roundabout} &   & \checkmark &  & & & \checkmark & & &  & &&&
 \\
 \cite{hildebrand2007unconventional} &   & \checkmark &  & & &  & & &  & &&&
 \\
 \cite{gallelli2008roundabout} &   & \checkmark &  &\checkmark  &\checkmark  &  & & &  &&&\checkmark& BOW
 \\
 \cite{jagannathan2004design} &   & \checkmark &  & &\checkmark &  & & &  & &&&
 \\
 \cite{koganti2012maximizing} &   & \checkmark &  & &\checkmark &  & & &  & &&&
 \\
 \cite{parsons2009parallel} &   & \checkmark &  & & &  & & &  \checkmark & &&&
 \\
 \cite{sayed2006upstream} &   & \checkmark &  & & &  & \checkmark & &  & &&&
 \\
 \cite{tabernero2006upstream} &   & \checkmark &  & &&  & \checkmark & &  & &&&
 \\
 \bottomrule
\end{tabular}
}
\label{table: uaidType}
\end{table}

With CORSIM and SimTraffic  being used as simulation platforms in some cases, the remaining studies used Vissim predominately. The only corridor-level study was conducted by \cite{Chou2014}, where a new UAID (named Tiangabout) was evaluated with a 15-intersection corridor in the State of Maryland. We also observed that the studies using empirical data \citep{yang2014development, Warchol2017, elazzony2010microsimulation, esawey2011unconventional,chilukuri2011diverging} are more likely to include the adjacent intersections in the simulation evaluation. The remainder of the studies focuses on isolated UAIDs. The majority of the studies made comparisons of a UAID with its conventional counterpart. For instance, a DDI is compared with a CDI; a MUT was compared with a standard four-leg signalized intersection. The roundabout-related geometric conditions (i.e., radius, island width, merging time gap, and lane width) are discussed by  \cite{gallelli2008roundabout}.

The use of measures depends on the prevailing traffic conditions. The number of stop may be suitable for a network that has not reached saturation, whereas total delay is apt for a network that is near its capacity (e.g.,a volume-to-capacity ratio greater than 0.8) \citep{Sen1997}.The most common measures for a signalized intersection are average delay per vehicle, numbers of stops, average queue length, and intersection throughput.  Among them, delay is the measure that relates driver's experience the most, as it represents the excess amount of time in traversing an intersection.  Delay can be further broken down to stopped time delay, approach delay, travel time delay, time-in-queue delay, and control delay \citep{Mathew2014}. Analytical delay prediction models (e.g., Webster's, Akcelik, HCM 2000) have been proposed along the years, but simulation provides an innovate and more robust and realistic way in evaluating the delays for intersections. 

The second common measure is the number of stops, which is an important parameter when it comes to air quality model, since the regaining of speed form a stopped vehicle requires additional acceleration, therefore burning more fuel. Queue length provides an indication of whether a given intersection impedes the vehicle discharging from an upstream intersection. Queue length is typically taken into account for corridor-level SPaT optimization. Other less common measures have been adopted for studies as well.  Hydrocarbon (HC), carbon monoxide (CO), nitric oxide (NO), and fuel consumption were used  by \cite{Chou2014} for evaluating the proposed Triangabout. The average speed and ratio of averaging moving time are used by \cite{hildebrand2007unconventional} and \cite{Reid2001}, respectively. The crash-related measures are lacking in the simulation studies because none of the traffic simulation software package is able to simulation collision to the best of our knowledge


Owing to the nature of the simulation, multiple scenarios could be tested systematically in an effective way. As shown in Table \ref{table: var_analysis}, three main categories of variables are observed in the previous studies: the geometric condition of the intersection, traffic flow characteristics, and SPaT plans for formulating multiple evaluation scenarios.  
Only two studies investigated the signalized crossover for DDI (\citep{Warchol2017}) and USC (\citep{sayed2006upstream}). The number of lane analysis tried to test the performance gain by each additional lane. Comparisons among multiple UAIDs were made for sensitivity analysis to investigate the comparative advantages of each UAID \cite{chlewicki2003new, hughes2010alternative, esawey2011unconventional,Reid2001, Tarko2017}. The majority of the studies researched the performance of UAIDs under various volume scenarios, some of which even exceed the capacity of a conventional intersection, to demonstrate the benefits of UAIDs. Heavy left-turn traffic and unbalanced split among intersection approaches were often combined with high overall volume to create the unbalanced and high demand scenarios where a conventional intersection typically reveals its inadequacy.

SPaT optimization depending on the traffic volume was performed in some studies to further improve the performance of UAIDs. Linear programming can be used for signal optimization \citep{koganti2012maximizing}, but commercial software specializing in SPaT are often chosen. Synchro was the most used software for SPaT optimization and it was adopted \citep{chlewicki2003new,hughes2010alternative, hildebrand2007unconventional, esawey2011unconventional, el2011operational,Park2017, autey2013operational, sayed2006upstream}. SIDRA, an alternative to Synchro,  was used by \cite{Chou2014} and \cite{elazzony2010microsimulation}. Recently, we have seen that the PTV Vistro has been used for signal optimization as well \citep{Warchol2017}. \cite{yang2014development} studied the optimal offset of the signalized crossover of the DDI.

The majority of the research showed the superior performance of UAIDs to conventional solutions. A DDI outperform a conventional diamond interchange if a location is with high traffic volume and with left-turn demand exceeding 50\% of the total demand \citep{Park2017}. It was concluded by \cite{Chlewicki2011} that the DDI could reduce 60\% of total intersection delay and 50\% of the total number of stops when designed properly.  The DLT intersection is able to potentially reduce average intersection delays in most traffic demand scenarios. A before-after study for the DLT at Baton Rouge, LA showed that the reduction in total crashes and fatality were 24\% and 19\%, respectively. The simulation also demonstrated 20\% to 50\% increase in throughput compared to a conventional intersection \citep{hughes2010alternative}. The reduction for a MUT in total crashes ranges from 20\% to 50\%\citep{scheuer1996evaluation, castronovo1995operational}.

\begin{table}[H]
\centering
\caption{Variables for Analysis}
\resizebox{1\textwidth}{!}
{
\begin{tabular}{p{0.8in}|p{1.5in}|p{4in}} \hline 
 & Variable/Factor & Study \\ \hline 
Geometric design\textbf{} & lane number & \cite{Tarko2017,Park2017,edara2005diverging,jagannathan2004design} \\ \hline 
\textbf{} & crossover spacing & \cite{Warchol2017,sayed2006upstream} \\ \hline 
\textbf{} & intersection type (for comparison) &\cite{chlewicki2003new,hughes2010alternative,
autey2013operational,elazzony2010microsimulation,Chou2014,Tarko2017,parsons2009parallel} \\ \hline 
Traffic\textbf{} & volume & \cite{Warchol2017,elazzony2010microsimulation,esawey2011unconventional,chilukuri2011diverging,
chlewicki2003new,hughes2010alternative,Park2017,autey2013operational,edara2005diverging,cheong2008comparison,
dhatrak2010performance,gallelli2008roundabout,hildebrand2007unconventional,jagannathan2004design,koganti2012maximizing,sayed2006upstream} \\ \hline 
\textbf{} & speed limit & \cite{gallelli2008roundabout} \\ \hline 
SPaT\textbf{} & phase configuration & \cite{Warchol2017,cheong2008comparison} \\ \hline 
\textbf{} & cycle length & \cite{esawey2011unconventional,koganti2012maximizing,cheong2008comparison} \\ \hline 
\textbf{} & signal offset & \cite{yang2014development} \\ \hline 
\end{tabular}
}
\label{table: var_analysis}
\end{table}

\subsection{Human Factor Study}

Unfamiliar urban intersections pose high cognitive demand on drivers who are prone to make unexpected maneuvers (e.g. hesitation, sudden stop, deviation from the planned path, sudden aggressive maneuvers \citep{autey2013operational,sayed2006upstream}. The driver's confusion was mentioned in most of the UAID studies as a potential drawback for UAIDs.  As we observed from practices, the off-ramp right tuning movements from the freeway in operation are signalized in DDIs due to the safety concern for unfamiliar drivers who may misidentify traffic on the opposite side of the roadway passing through a DDI interchange \citep{chilukuri2011diverging}. \cite{Reid2001} stated the reduction in delay and travel time would be less after accounting for driver's confusion.  

Traditionally, human factor study for intersection was carried out by conducting a study of empirical data \citep{petridou2000human}, or survey of road users \citep{gkartzonikas2019have}.  In recent years, the advancement of computational power makes driving simulator more prevalent than ever before. A driving simulator constructs a realistic virtual environment where a driver can be ``placed'' into the network for driving tasks.  Using driving simulator for assessing the driver's confusion is virtually risk-free. It is also very cost-effective to rapidly and systematically test a wide range of scenarios (different geometric configurations, pavement markings, and signage). 

The FHWA, in collaboration with Missouri DOT, evaluated the human factor of DDI using the Highway Driving Simulator located in the Turner-Fairbank Highway Research Center.  A DDI and a CDI were simulated. Seventy-four drivers within the Washington D.C. area were recruited for the experiment which aimed to investigate the wrong way violation, navigation errors, red-light violations, and driving speed through the DDI \citep{Bared2007}. \cite{zhao2015driving} found that wrong way crashes inside the crossroad between ramp terminals accounted for 4.8\% of the fatal and injury crashes occurring at the DDI. 

\cite{zhao2015driving} studied the human factor aspect of the exit-lanes for left-turn (EFL) intersection by using a driving simulator.  Being classified as a UAID, the EFL opens an exit-lane at the opposing travel lane to be used by left-turn traffic with an additional traffic light installed at the median opening. The possible confusion for the new users, the adequacy of signage marking, and the cue movement from by other vehicles were investigated. Navigation errors, utilization of the mixed-usage-area, red-light violation, and wrong-way violation data were collected from 64 participants. The results showed that it was dif?cult for participants to comprehend EFL's special operation procedure without prior training or experience. It took an average of five seconds longer for the unfamiliar driver to use mixed-use-area of an EFL on the opposing travel lane.  \citep{tawari2016predicting} proposed a prediction framework for unexpected behaviors as a precursor for accident intervention based on three sources of data: vehicle dynamics, visual scanning of drivers, and the difference in vehicle dynamics between infrastructure expectation and the actual ones.

\subsection{Challenge for UAID}
Unfortunately, the larger footprint for UAIDs in some cases makes them infeasible to be deployed in an urban environment. For instance, the DLT intersection, when implemented on four approaches, has a footprint that is nearly one acre larger than a conventional intersection \citep{hughes2010alternative}. The cost of ROW acquisition is the generally the greatest cost of constructing a UAID and it varies depending on locales.  Nearly all UAIDs cost more than the conventional intersection with the exception of the Jughandles \citep{hildebrand2007unconventional}.  The complex re-routing of traffic may also be challenging to comprehensive for unfamiliar drivers. The unfamiliarity to the UAID is the primary reason for the navigation error, which is deemed as the wrong-way maneuvers when it comes to driving through a UAID. Lastly, the secondary intersections of a UAID could potentially create delays that impact the overall traffic \citep{li2018symmetric}.

\section{Signalized UAID with CAV}
\label{sect: signalizedCAV}
Information exchanged between vehicles and the signal controller had been scarce prior to CAV. Traditionally, the intersection controller detects the presence of vehicles via cameras or loop detectors. The SPaT plan was adjusted based on the available information, then vehicles adjust driving accordingly to the signal phases as they approach the intersection. CAV is promised to enhance the signal control paradigm in a significant way with real-time information transmitted via wireless communication. 
The connectivity of CAVs is able to obtain critical information regarding infrastructure (e.g., SPaT plan, UAID information); the automation can improve the trajectory for the approaching vehicles. Specifically for UAIDs, ADAS can reduce driver's confusion by providing navigational information or even actively prevent the driver from performing incorrect maneuver (making a wrong turn onto the travel lane of the opposing traffic). 



\subsection{Improvement on SPaT}

ATCSs is a good fit for coordination in a traffic-responsive manner. However lack of reliably traffic information remains one of the limiting factors for ATCS. Fixed sensor, such as magnetic detectors, can only collect traffic flow information at discrete locations. Study the utilization of the high-resolution real-time vehicle information data from CAV for ATSCs is an ongoing effort. A bi-level optimal ATCS algorithm, which used real-time data from CAV via V2I communication, was proposed by \cite{feng2015real}. An estimation algorithm for HVs was also developed to construct a complete arrival table of all vehicles for SPaT allocation. 
Studies that demonstrated the potential improvements because of  real-time traffic information of CAVs was also reported by \cite{goodall2014microscopic,He2012,priemer2009decentralized}. 

The benefits brought to ATSCs extends to UAIDs. The geometry of UAIDs typically requires more signalization but with each controller with fewer phases due to the reduced movements. The coordination among closely-spaced signal controllers, as a result, is an important task for the performance of UAIDs. For example, the DDI and DLT both require an up-stream crossover phases.  Thus far, the coordination in UAID is using pre-timed signal that maximized the green band of movement with high priority, therefore ensuring good traffic progression. 

\subsection{Vehicle Trajectory Optimization}

\paragraph*{Eco-driving}
is another widely-researched direction for increasing intersection performance.  The Eco-driving guidance system aims to increase the fuel efficiency by minimizing the acceleration, deceleration, and idling with an optimal speed profile with SPaT information. \cite{jiang2017eco} proposed an eco-driving system that prioritized mobility, instead of fuel efficiency, in improving the overall traffic flow. It was found that the reduction in fuel consumption and CO2 emission are up to 58\% and 33\%, respectively. Additionally, the throughput was increased by approximately 11\%.	\cite{zohdy2012intersection} proposed an intersection management system focusing on optimizing vehicle arrival.  The system assumes 100\% penetration of CAVs with perfect V2I communication and it was tested with a hypothetical standard four-leg signalized intersection. 

\paragraph*{Platooning}
has been used to access the traffic signal progression at a corridor level in SPaT optimization \citep{todd1988effects}. Platoon in CAV refers to a group of consecutive closely-coupled vehicle string. Platoon-based control is only technologically possible with the introduction of CAV. Vehicles and platoons are able to be identified, so the platoon spiting and merging are possible, hence the platoon-based SPaT. 
	 
\paragraph*{Queue Discharging} aspect of signalized intersection can be improved by CAV. \cite{Zhong2017a} studied a CAV-based application on real-world signalized intersection using Vissim. The start-up lost time was assumed to be zero owing to V2X communication and all the CAVs within a platoon moved synchronously upon the commencement of a green phase. Without changing the existing SPaT plan, the average stop delay was reduced by 17\% when the MPR of CAV reached 70\%. \cite{le2016automated} studied the queue discharging operation of CAVs with assured-clear-distance-ahead principle by using a deterministic simulation model. They observed, on the contrary to the study by \cite{Zhong2017a}, only marginal improvement to intersection throughput due to the synchronous start-up movement.  However, they found that the processing time for a ten-vehicle queue did reduce by 25\% with full CAVs, compared to that for the HV-queue with the same amount of vehicles.

\paragraph*{Operation with Non-equipped Vehicles} is an unavoidable stage in the near-term deployment of CAV.
Researchers have been studying the possible cooperative  between CAVs and HVs by strategically consider the following HVs for intersection \citep{zhao2018platoon}. \cite{yang2016isolated} outlined their bi-level optimal intersection control algorithm that was able to factor in the trajectory design for CAVs, and the prediction of HVs based on real-time CAV data. The prediction of the trajectory of HVs is based on Newell's car-following model and the positional information of CAVs. The baseline used for comparison was an actuated signal control algorithm under a range of traffic demand between 1000 and 2000 vehicle per hour (vph).


\subsection{Virtual Traffic Light}

The virtual traffic light (VTL) was initially proposed as a  self-organized intersection traffic control based on only V2V communication \citep{Ferreira2010}.  Under VTL, an elected vehicle acts as the temporary road junction infrastructure to broadcast traffic light message when it approaches an intersection. The other drivers in the adjacent area are conveyed with the crossing message via in-vehicle display. The selection of a leader in VTL is referred as leader election protocol (LEP). \cite{Fathollahnejad2013} investigated the probability of disagreement among participating vehicles in LEP and presented a series of simple round-based consensus algorithms for solving the selection problem. The number of participating vehicles, the rounds of message exchange, the probability of message loss, and the decision criterion (assume or abort as a leader for a vehicle) were taken into account.  However, the feasibility of designing a consensus algorithm that ensures safety under asymmetric information (i.e. the number of candidates is unknown.) remained an open question. \cite{sommer2014networking} adapted the dynamic ad hoc network algorithm developed by \cite{Vasudevan2004} as the LEP and tested the network with Veins (a simulation environment comprised of OMNeT++ and SUMO) on a real-world network.  An extension of the VTL algorithm was developed in subsequent research by the Sommer's group for managing arbitrary intersection geometries \citep{Hagenauer2014}.


The LEP process, which ensures one and only one leader is selected, remains a major barrier for its implementation. The secondary intersections in UAID require multiple LEP processes, which will further increase the computational expense. Furthermore, VTL also requires symmetric information among the approaching vehicles (i.e. no unknown agents). Hence, the VTL is incompatible with non-CAV vehicles as they are not able to participate in the LEP process.   Moreover, the information exchanged among all vehicles could pose a heavy burden on the wireless communication network. As such, VTL is only feasible under high MPR of CAVs when the communication technology is sufficient. Additionally, the large footprint of UAID may cause issues with the communication aspect of the VTL.


\section{Signal-free UAID with CAV}
\label{sect: signalFreeCAV}
The signal-free autonomous intersection control (AIM) is the re-imaging of intersection control with the new level of automation and connectivity owing to CAV technology.  There are key components of AIM: reservation system, priority policy, and vehicle control. This section focus on connected the key concept of AIM to UAID without the loss of generality.

\subsection{Reservation System}
Compared to a signalized intersection, the main differences are 1) the conflict separation is made at vehicle level; 2) the priority sequence need not follow the pre-determined sequence as in signalization; 3) the switching among permission to enter can occur more rapidly due to automation.

Thus far, there are four types of reservation systems: intersection-based reservation, tile-based reservation, conflict point-based reservation, and vehicle-based reservation. Intersection-based reservation \citep{Bashiri2017, carlino2013auction, jiang2017distributed, Carlson2016, du2018hierarchical} only allow one vehicle within an intersection at any given time. Tile-based reservation system \citep{Bichiou2018, Zohdy2012,wuthishuwong2015safe, liu2019trajectory} discretizes the space into a grid, the unavailability of a reservation is confirmed by the overlapping tile with existing reservations. The conflict point-based reservation creates the separation at the conflict points where collision occur potentially \citep{kamal2015vehicle, Ding2017a, Fajardo2011, levin2017conflict}. The last type of reservation is vehicle-based reservation where the trajectory of vehicles are unlimited \citep{Li2018, Li2018a}.


For UAIDs, the reservation of the entire intersection is rather counterproductive, as the footprint of UAIDs is not trivial.  Take DDI as an example, its merging zone becomes a rectangular shape with the long edge spans 300-meter long. Owing to the movement separation geometric characteristics of UAID, the conflict point-based and tile-based reservation system are more suitable. However, the increased computational expenses for such two types of reservations need to be taken into account. The vehicle-based reservation system is expected to have the same level of computational intensity, if not more, of the standard intersection, which is not near to real-time implementation.


\subsection{Trajectory Planning}
\begin{figure}[H]
	\centering
	\includegraphics[width=0.7\columnwidth]{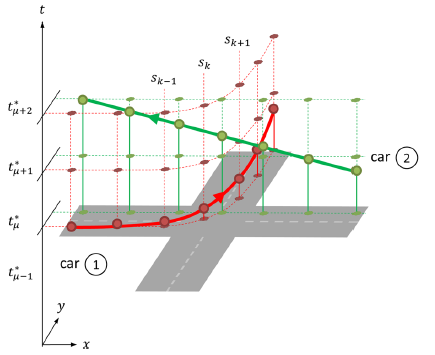}
	\caption{Vehicle-level conflict avoidance \citep{wuthishuwong2015safe}} 
\label{fig: trajPnanning}
\end{figure}

As mentioned, the permission to enter an intersection for AIM is considered at a vehicle level, compared to signalization which works at a vehicle group level. As demonstrated in Figure \ref{fig: trajPnanning}, vehicle trajectory planning has been used to separate conflicts, except in the intersection-based reservation.

Centralized vehicle trajectory planning has been adopted in many AIMs, in which the intersection menager (presumably hosted on RSU) controls the CAVs. \cite{lee2012development} proposed a cooperative vehicle intersection control (CVIC) framework which did not require any traffic signal even under moderate intersection demand (1900 vph).  In CVIC, each vehicle was assigned with an individual trajectory via V2I communication. \cite{kamal2015vehicle} proposed a signal-free, model predictive control framework-based intersection control framework that globally coordinates all vehicles with optimal trajectories which were computed based on the avoidance of cross-collision risks.  An optimal intersection control system, which is designed to minimize travel time for CAV was proposed by \cite{Bichiou2018}. Two versions of the proposed framework-the optimal control time and the optimal control effort-were tested on a roundabout, an all-way-stop-controlled intersection, and a signalized intersection.  A significant reduction in CO2 emission was observed, however the proposed model has high computational cost from conducting nonlinear optimization; it takes up to five minutes to solve the optimization for a set of four vehicles. A centralized cooperative intersection control was proposed by \cite{Ding2017a} where the control strategy was  to minimize intersection delay, fuel consumption, and emission.

Centralized intersection management strategies are costly to implement and their scalability is open to question, even under the assumption that a central controller has the ability to track and schedule hundreds or even thousands of vehicles in real time.  Additionally, the current state of V2X wireless communication cannot technologically guarantee such performance with thousands of vehicles in the vicinity of an intersection. 

Decentralized intersection management relies on local information among vehicles \citep{makarem2011decentralized}. A fully-distributed heuristic intersection control strategy was proposed by \cite{hassan2014fully} with the objective to minimize the information being exchanged in each time step. Here, the vehicles  are categorized into four groups: ``Out'', ``Last", ``Mid'', and ``Head''. The ``Head'' group is closest to the intersection and it assumes the role of the intersection manager. This management strategy was subsequently enhanced for real-time application by \cite{Bichiou2018b}, in which the initial nonlinear constraints \citep{hassan2014fully} was simplified.  A decentralized intersection navigation faces two main challenges: predefined travel paths, and the possible local minimums that lead to intersection deadlock \citep{makarem2012fluent}. 
	
Game theory can be used to determine the priority among a group of vehicles without the need for signal control. Cooperative games can be organized among vehicles and the manager via V2X communication. \cite{Zohdy2012} proposed a CACC-CG (Cooperative Adaptive Cruise Control – Cooperative Game) schema which is consisted of a manager agent (RSU) and reactive agents (CAVs). With the symmetrical information shared among CAVs, the reactive agents choose among three actions: acceleration, deceleration, or maintaining current speed. The minimum utility value from the payoff table was chosen by each CAV at each time step. The CACC-CG control reduced 65\% intersection delay in comparison with a four-way-stop-sign control. 
\cite{Elhenawy2015} proposed a signal-free intersection management where CACC-equipped vehicles communicate vehicle status to a centralized intersection controller. With the vehicle information, the controller solves the game matrix and obtains the Nash equilibrium. Then the optimal action was distributed to each vehicle. Compared to an all-way stop control intersection, the proposed scheme achieved 49\% and 89\% reduction in vehicle travel time and delay, respectively. 
 

\subsection{Limitation of Signal-free Intersection Control}

There are drawbacks for AIM at the current stage. The reservation-based intersection control methods can disrupt platoon progressive, causing greater delays than conventional traffic signals \citep{levin2016paradoxes}. 
According to the Highway Capacity Manual \citep{manual2010highway},  the saturation flow rate of a signalized intersection is within the range of 1700-1900 vph. \cite{levin2016paradoxes}, using a probabilistic model, computed the saturation flow rate of a conflict-point-based, signal-free, standard intersection without turning movement to be 1667 vph.  \cite{zhang2018penetration} compared the optimal control non-signalized intersection control with signalized control and concluded that as traffic grows, a higher CAV penetration is necessary to match the performance of signalization. When the traffic demand is above the critical flow rate (750 veh/hr-ln), even 100\% CAV penetration still cannot outperform signalization in terms of energy saving. Under the saturated condition, nearly all vehicles have to slow down or even stop to create the necessary separation for entering the intersection. However, the signal-free AIM is more effective for reducing travel delay than signalization.

	The computation complexity of AIM remains as one of the major issues for real-time implementation. System-optimal trajectory planning is often formulated as nonlinear programming problem. Not to mention the bandwidth required between vehicles and intersection manager.  Additionally, the nature of the UAID adds complexity to the reservation system. For instance, the opposing through movements of a DDI has two conflict points whereas there is none for a standard intersection.


\section{Discussion and Future Research}
\label{sect: Discussion}

Intersections remain the common traffic bottlenecks in the modern transportation network. UAID has shown promising benefits in its initial deployments. It is believe that the performance of UAIDs can be further improved by CAV technology. This section aims to provide discussion for future research for UAID.

\subsection{Signal-Free Intersection Management}

The signal-free intersection control paradigm aims to eliminate the traditional signals and coordinate crossing at a vehicle level. Nearly all of the proposed signal-free control schemes are still under the concept development stage.  First, they require homogeneous CAV traffic in order to successfully operate. Second, more experiments accounting for a wide spectrum of scenarios are needed. Insofar, the testing scenarios are overly simplified. For instance, a standard four-leg intersection with only through movements and four one-lane approaches is far from the real-world condition. Third, its good performance under unrealistically low traffic volume is not guaranteed in medium or high volume scenarios.  When the traffic volume increases to the magnitude of thousands of vehicles per hour, a scale that is typically in the real-world deployment, the computational complexity of these control algorithm is unlikely to scale linearly. To the best of our knowledge, no computational deadline was set for the scheduling processing among the reported studies, which means the controller was allowed to take as much time for sorting out the enter sequence. This is very unlikely the case for real-world deployment.

\subsection{SPaT Coordination}
Little information regarding possible coordination with adjacent signalized intersections for UAID is available.  Driving behaviors were not calibrated in previous UAID studies. At a corridor level, the impact of driving behavior may be more pronounced. Failing to take it into account could affect the simulation results. 

Most of the evaluation of UAID is performed via simulation. Validation of the simulation result for UAID is also needed. Comparisons should be made between the simulation models, which were used for UAID development, and the actual performance of the UAID in operation.  The difference, if any, could reveal the factors that were not initially considered in the simulation models, such as the driver's confusion. The potential confusion for drivers when facing UAIDs has been one of the primary concerns for UAIDs, which has not been adequately assessed in the previous research. For the majority of the simulation studies, the driver confusion has been assumed non-existent. The performance gain for UAID could potentially be inflated if the impact of the confused drivers is proven to be significant.  Previous research rarely evaluated the performance of UAID at a corridor level with few exceptions 

The CAV technology could be an excellent complement for the UAIDs. The connected environment, enabled by the wireless communication between a vehicle and the intersection infrastructure, is able to provide geometry information to help unfamiliar drivers to navigate through UAIDs. The potential aid gained from CAV technology, in the near-term could improve the performance of UAID by abating or even eliminating driver's confusion.

\subsection{New UAID Designs }
CAV warrants more versatile UAIDs for signalized intersections. \cite{sun2017capacity} proposed an intersection management, called Maximum Capacity Intersection Operation Scheme with Signals (MCross). The operation of MCross intersection utilizes all the travel lanes at any time. The lane assignment, which is obtained by solving the mixed-integer non-linear multi-objective optimization, relied entirely upon fully-automated CAVs. Li proposed a new UAID, called Symmetric intersection \citep{li2018symmetric}, which only needs three phases to separate all conflicts by weaving left-hand driving rule and right-hand driving rule. He used a capacity maximization analytical model to prove the effectiveness of the proposed UAID.  However, such innovative intersection design could have serious ramification if navigation error occurs. Considering the complexity of the intersection operation of the Symmetric intersection, the transition between left-hand driving and right-hand driving rule for human driver could cause confusion and leave little margin for error. In the worst-case scenario, a driver could end up on a travel lane of the opposing traffic.

\subsection{Mixed Traffic Operation}

The Volpe National Transportation System Center estimated that it might take 25 -30 years for CAVs to reach a 95\% MPR, even with a federal mandatory installation of DSRC devices on new light vehicles manufactured in the US \citep{volpe2008vehicle}. As promising as they may sound for these signal-free intersections, their requirement of fully-automated vehicles (SAE Level 5) renders them infeasible in the near future, where CAVs are deployed in mixed traffic conditions. Therefore compromise for signal-free intersection management has to be made. A signal-free framework that is compatible with human operation was proposed and tested in \citep{dresner2007sharing}. The performance of FCFS-based autonomous intersecton management degraded significance with 5-10\% presence of human operation due to traffic light control for human operation and its resulting blockage to automated vehicle. An enhanced version was proposed by \cite{stone2015autonomous}, where the control of a vehicle can be provisionally transferred to the automated driving system. The traffic signal was employed as a fallback strategy when a reservation cannot be obtained. Under this hybrid framework, the similar performance for the original design can be achieved with no more than 40\% MP. However, additional research is still required to quantify the possible trade-off that has to make when it comes to semi-AIM.

	Using CAV vehicle to influence or indirectly control non-equipped vehicle is another popular direction for dealing with heterogeneous traffic. This has been tested in CAV-based speed harmonization application \citep{malikopoulos2018optimal}. Using CAV as pace car to optimized intersection discharging pattern have been reported \citep{jiang2017eco, gong2018cooperative}.  The issue contains two main parts. The first is the model that mimics the car-following behavior of a human-driven vehicle with a high degree of accuracy. The second problem  pertains to how a CAV reacts to the resulting behavior of the non-equipped vehicle behind it. Either of them has been actively studied.

\subsection{Human Factor}

The long-term impacts of vehicle automation to human drivers have yet to been seen.  Potential negative impacts include overreliance to the ADAS, erratic mental workload, driving skill degradation, and decline situation awareness \citep{Saffarian2012}. Driving simulators have been used to study the driver's behaviors for navigating UAIDs. The impact of geometric configurations of UAIDs, effective pavement markings, and signage, which provides navigational cues and helps to alleviate driver's confusion, were the focuses of the previous simulation studies. Further evaluations are still needed. Ideally, such driving behavior study should be conducted at the initial design stage of a new UAID, as done by \cite{zhao2015driving}.  Given the access to a driving simulator has kept increasing in the past decade, high-fidelity driving simulators that provide an immersing driving experience for participants remain costly and are only accessible to limited entities (e.g., the Turner-Fairbank Highway Research Center). Additionally, the integration of a driving simulator and a traffic simulator (e.g., Vissim, Aimsun) that provides a more realistic representation of prevailing traffic conditions is gaining popularity in the research community.  Such integration bridges the individual-level driving behavior and the network-level traffic flow.

The human-machine interface (HMI) of ADAS systems for drivers is another crucial area that is worth exploring.  How to convey actionable information or instructions to human drivers is an emerging topic. 
As studies show, feedback provides prematurely or needlessly frequently could result in distractions or even dismissal of the system entirely by drivers \citep{Saffarian2012}. Furthermore, how the transfer of responsibility (authority) between human to the ADAS should be performed in some hybrid signal-free intersection \citep{stone2015autonomous}? These are few of the open questions yet to be answered concerning ADAS. 

From a technological standpoint, the determination of the erratic or potential danger maneuvers from confused drivers is vital in the deployment of ADASs in eliminating the impact of driver's confusion. Assuming the destination of a vehicle is known by the RSU in an anonymized manner, the prediction framework for unexpected behaviors is instrumental in the successful intervention of ADASs for confused drivers. For instance, the difference in vehicle dynamics from the expectation of the infrastructure (e.g., RSUs) and the actual vehicle dynamics, combined with the visual scanning of drivers, were used to detect the unanticipated behaviors in \citep{tawari2016predicting}.  

The concern for driver's confusion is one of the barriers for adoptions of UAIDs, but it has not been taken into account in the previous studies.

\section{Conclusion}
\label{sect: conclusion}
Intersections have been one of the main sources for control delay and severe collisions.  Researchers and engineers have conceptualized and deployed a series of innovative unconventional arterial intersection designs (UAIDs) which alters the nature and the number of the conflict points of a conventional intersection. UAIDs have started to gain recognition in improving mobility and safety of intersections because of the reduced signal phases and conflict points.  In this paper,  we systematically review the state of the research of UAID. We then discuss benefits that CAV bring to UAIDs for signalized and signal-free paradigms. With connectivity and automation, CAV is believed to be a great complement for UAIDs. The highlights of the trend and future research topics related to UAID with CAV are provided in the end.

\bibliographystyle{apacite}
\bibliography{uaid}

\end{document}